\documentclass{PoS}

\title{Neutrinos versus Quarks, MNS versus CKM}

\ShortTitle{Neutrinos vs. Quarks, MNS vs. CKM}

\author{\speaker{Mu-Chun Chen}\\
        Department of Physics \& Astronomy, University of California, Irvine, CA 92697-4575\\
        E-mail: \email{muchunc@uci.edu}}

\author{K.T. Mahanthappa\\
        Department of Physics, University of Colorado, Boulder, CO 80309-0390\\
        E-mail: \email{ktm@pizero.colorado.edu}}

\abstract{We review recent developments in models of fermion masses and mixing for both  quark and lepton sectors. Emphases are given to models based on finite group family symmetries. In particular, we describe one recent model based on SU(5) combined with a family symmetry based on the double tetrahedral group, $T^{\prime}$. In this model, a near tri-bimaximal MNS matrix and a realistic CKM matrix are simultaneously generated;  the MNS matrix gets  slightly modified by virtue of having the Georgi-Jarlskog relations. Due to the presence of complex Clebsch-Gordon coefficients in $T^{\prime}$, CP violation in this model is entirely geometrical in origin.}

\FullConference{The Sixth International Conference on Flavor Physics and CP Violation (FPCP2009)\\
		 May 27 - June 1, 2009\\
		 Lake Placid, New York}

\begin{document}

\section{Introduction}

One of the most significant unsolved questions in particle physics is the origin of fermion mass hierarchy and flavor mixing. Even though the Standard Model (SM) works beautifully in explaining all particle interactions, it has many free parameters in the Yukawa sector that accommodate the observed masses and mixing angles for quarks and leptons. The  number of free parameters can be greatly reduced by expanding the SM gauge group to a grand unified gauge symmetry, which gives rise to inter-family relations that connect quarks and leptons within the same family. Further reduction of parameters can be achieved with an additional family symmetry that relates quarks and leptons of different families. (For reviews, see {\it e.g.} \cite{Chen:2003zv,Chen:2008eq}.)

The recent  advent of the neutrino oscillation data from Super-Kamiokande gives further support to models based on Grand Unified Theories (GUTs), in which the seesaw mechanism can arise  naturally. The global fit to current data from neutrino oscillation experiments give the following best fit values and $2\sigma$ limits for the mixing parameters~\cite{Maltoni:2004ei},
\begin{eqnarray}
\sin^{2} \theta_{12} & = & 0.30 \; (0.25 - 0.34),  \nonumber \\
\sin^{2} \theta_{23} & = &  0.5 \; (0.38 - 0.64), \nonumber \\
\sin^{2} \theta_{13} & = & 0 \;  (< 0.028), \nonumber \\
\Delta m_{12}^{2} & = & 7.9 \; (7.3 - 8.5) \; \mbox{eV}^2, \nonumber\\
\Delta m_{23}^{2} & = & 2.2 \; (1.7 - 2.9) \; \mbox{eV}^2. \nonumber
\end{eqnarray}
In addition, recent analyses~\cite{Fogli:2008jx} from the Bari group have given hints on possible non-zero value for $\theta_{13}$, with 
\begin{displaymath}
\sin^{2}\theta_{13} = 0.016 \pm 0.010 \; , 
\end{displaymath}
at $1 \, \sigma$.  These experimental best fit values for the mixing parameters are very close to the values arising from the so-called ``tri-bimaximal'' mixing (TBM) matrix~\cite{Harrison:2002er},
\begin{equation}
U_{\mathrm{TBM}} = \left(\begin{array}{ccc}
\sqrt{2/3} & 1/\sqrt{3} & 0\\
-\sqrt{1/6} & 1/\sqrt{3} & -1/\sqrt{2}\\
-\sqrt{1/6} & 1/\sqrt{3} & 1/\sqrt{2}
\end{array}\right) \; , \label{eq:tri-bi}
\end{equation}
which predicts 
\begin{eqnarray}
\sin^{2}\theta_{\mathrm{atm, \, TBM}} & = & 1/2 \; ,  \nonumber\\
\sin^{2}\theta_{\odot, \mathrm{TBM}} & = & 1/3 \; , \nonumber\\
\sin\theta_{13, \mathrm{TBM}} & = & 0\; .
 \end{eqnarray}
Even though the predicted $\theta_{\odot, \mathrm{TBM}}$ is currently still  allowed by the experimental data at $2\sigma$, as it is very close to the upper bound at the $2\sigma$ limit, it  may be ruled out once more precise measurements are made in the  upcoming experiments.  

One of the challenges that GUT models face is to give rise to large neutrino mixing and at the same time accommodate small quark CKM mixing. It has been pointed out that the tri-bimaximal mixing matrix can arise from a family symmetry in the lepton sector based on $A_{4}$~\cite{Ma:2001dn}.  However, due to its lack of doublet representations, CKM matrix is an identity in most $A_{4}$ models. It is hence not easy to implement $A_{4}$ as a family symmetry for both quarks and leptons~\cite{Ma:2006sk}.

\section{The Model}

In \cite{Chen:2007afa,Chen:2009gf}, a grand unified model based on SU(5) combined with the double tetrahedral group~\cite{Frampton:1994rk}, $T^{\prime}$, was constructed, which successfully gives rise to near tri-bimaximal leptonic mixing as well as realistic CKM matrix elements for the quarks. The group $T^{\prime}$ is the double covering group of $A_{4}$. In addition to the $1, \; 1^{\prime}, \; 1^{\prime\prime}$ and $3$ representations that $A_{4}$ has, the group $T^{\prime}$ also has three in-equivalent doublet representations, $2, \; 2^{\prime}, \; 2^{\prime\prime}$. This enables the $(1+2)$ assignments, which has been shown to give realistic masses and mixing pattern in the quark sector~\cite{so10ref}. 

One special property of $T^{\prime}$ is the fact that its Clebsch-Gordon coefficients are intrinsically complex, independent of the basis for the two group generators. This thus affords the possibility that CP violation can be entirely geometrical in origin~\cite{Chen:2009gf}.  (We note that in addition to the capability of giving rise to mixing angles and CP violation from CG coefficients, the group $T^{\prime}$ has recently been utilized in a Randall-Sundrum  model to avoid tree-level flavor-changing neutral currents~\cite{Chen:2009gy}, which are present in generic RS models.)

The charge assignments of various fields in our model are summarized in Table~\ref{tbl:charge}.  Due to the transformation properties of various fields, only top quark mass is allowed by the $T^{\prime}$ symmetry, and thus it is the only mass term that can be generated at the renormalizable level. To give masses to the lighter generations of fermions, which transform non-trivially under $T^{\prime}$, the $T^{\prime}$ symmetry has to be broken, which is achieved by a set of flavon fields. 
\begin{table*}[h!]
\caption{Charge assignments. Here the parameter $\omega = e^{i\pi/6}$.}  
\label{tbl:charge} 
\vspace{0.06in}
\begin{tabular}{|c|ccc|ccc|cccccc|cc|}\hline
& $T_{3}$ & $T_{a}$ & $\overline{F}$ & $H_{5}$ & $H_{\overline{5}}^{\prime}$ & $\Delta_{45}$ & $\phi$ & $\phi^{\prime}$ & $\psi$ & $\psi^{\prime}$ & $\zeta$ & $N$ & $\xi$ & $\eta$  \\ [0.3em] \hline\hline
SU(5) & 10 & 10 & $\overline{5}$ & 5 &  $\overline{5}$ & 45 & 1 & 1 & 1 & 1& 1 & 1 & 1 & 1\\ \hline
$T^{\prime}$ & 1 & $2$ & 3 & 1 & 1 & $1^{\prime}$ & 3 & 3 & $2^{\prime}$ & $2$ & $1^{\prime\prime}$ & $1^{\prime}$ & 3 & 1 \\ [0.2em] \hline
$Z_{12}$ & $\omega^{5}$ & $\omega^{2}$ & $\omega^{5}$ & $\omega^{2}$ & $\omega^{2}$ & $\omega^{5}$ & $\omega^{3}$ & $\omega^{2}$ & $\omega^{6}$ & $\omega^{9}$ & $\omega^{9}$ 
& $\omega^{3}$ & $\omega^{10}$ & $\omega^{10}$ \\ [0.2em] \hline
$Z_{12}^{\prime}$ & $\omega$ & $\omega^{4}$ & $\omega^{8}$ & $\omega^{10}$ & $\omega^{10}$ & $\omega^{3}$ & $\omega^{3}$ & $\omega^{6}$ & $\omega^{7}$ & $\omega^{8}$ & $\omega^{2}$ & $\omega^{11}$ & 1 & $1$ 
\\ \hline   
\end{tabular}
\vspace{0.1in}
\end{table*}

Due to the presence of the  
$Z_{12} \times Z_{12}^{\prime}$ symmetry, only nine operators are allowed in the model, and hence the model is very predictive, the total number of parameters being nine in the Yukawa sector for the charged fermions and the neutrinos. The Lagrangian of the model is given as follows,
\begin{eqnarray}
\mathcal{L}_{\mathrm{Yuk}} &  = &   \mathcal{L}_{\mathrm{TT}} + \mathcal{L}_{\mathrm{TF}} + \mathcal{L}_{\mathrm{FF}} \; \\
\mathcal{L}_{\mathrm{TT}} &= & y_{t} H_{5} T_{3}T_{3} + \frac{1}{\Lambda^{2}} y_{ts} H_{5} T_{3} T_{a} \psi \zeta 
+ \frac{1}{\Lambda^{2}} y_{c} H_{5} T_{a} T_{a} \phi^{2} + \frac{1}{\Lambda^{3}} y_{u} H_{5} T_{a} T_{a} \phi^{\prime 3}   \; , \\  
\mathcal{L}_{\mathrm{TF}}  &=&  \frac{1}{ \Lambda^{2}}  y_{b} H_{\overline{5}}^{\prime} \overline{F} T_{3} \phi \zeta 
 + \frac{1}{\Lambda^{3}} \biggl[ y_{s} \Delta_{45} \overline{F} T_{a} \phi \psi N  + 
 y_{d} H_{\overline{5}}^{\prime} \overline{F} T_{a} \phi^{2} \psi^{\prime}  \biggr] \; , \\
\mathcal{L}_{\mathrm{FF}} & =&   \frac{1}{M_{x}\Lambda} \biggl[\lambda_{1} H_{5} H_{5} \overline{F} \, \overline{F} \xi +  \lambda_{2} H_{5} H_{5} \overline{F} \, \overline{F} \eta\biggr] \; , 
 \end{eqnarray}
 where $M_{x}$ is the cutoff scale at which the lepton number violation operator $HH\overline{F}\, \overline{F}$ is generated, while $\Lambda$ is the cutoff scale, above which the $T^{\prime}$ symmetry is exact.  (For the VEV's of various scalar fields, see Ref.~\cite{Chen:2007afa}.) The parameters $y$'s and $\lambda$'s are the coupling constants.

The interactions in $\mathcal{L}_{TT}$ and $\mathcal{L}_{TF}$ gives rise to the up-type quark and down-type quark  mass matrices, $M_{u}$ and $M_{d}$, respectively. Since the lepton doublets and iso-singlet down-type quarks are unified into a $\overline{5}$ of $SU(5)$, their mass matrices are related. Upon the breaking of $T^{\prime}$ and the electroweak symmetry, these mass matrices are given in terms of seven parameters by~\cite{Chen:2007afa}
\begin{eqnarray}
M_{u} & = & \left( \begin{array}{ccc}
i \phi^{\prime 3}_{0}  & (\frac{1-i}{2}) \phi_{0}^{\prime 3} & 0 \\
(\frac{1-i}{2})  \phi_{0}^{\prime 3}  & \phi_{0}^{\prime 3} + (1 - \frac{i}{2}) \phi_{0}^{2} & y^{\prime} \psi_{0} \zeta_{0} \\
0 & y^{\prime} \psi_{0} \zeta_{0} & 1
\end{array} \right) y_{t}v_{u}, \qquad \\
M_{d}  & = & \left( \begin{array}{ccc}
0 & (1+i) \phi_{0} \psi^{\prime}_{0} & 0 \\
-(1-i) \phi_{0} \psi^{\prime}_{0} & \psi_{0} N_{0} & 0 \\
\phi_{0} \psi^{\prime}_{0} & \phi_{0} \psi^{\prime}_{0} & \zeta_{0} 
\end{array}\right) y_{d} v_{d} \phi_{0} \; , \\
M_{e} & = & \left( \begin{array}{ccc}
0 & -(1-i) \phi_{0} \psi^{\prime}_{0} & \phi_{0} \psi^{\prime}_{0} \\
(1+i) \phi_{0} \psi^{\prime}_{0} & -3 \psi_{0} N_{0} & \phi_{0} \psi^{\prime}_{0} \\
0 & 0 & \zeta_{0} 
\end{array}\right) y_{d} v_{d} \phi_{0} \; . 
\end{eqnarray}
which manifest the $SU(5)$ relation, 
\begin{equation}
M_{d} = M_{e}^{T} \; ,
\end{equation} 
except for the factor of $-3$ in the (22) entry of $M_{e}$, due to the $SU(5)$ CG coefficient through the coupling to $\Delta_{45}$. In addition to this $-3$ factor, the Georgi-Jarlskog (GJ) relations at the GUT scale,
\begin{eqnarray}
m_{e} & \simeq & \frac{1}{3} m_{d} \; , \\
 \quad m_{\mu} & \simeq & 3 m_{s} \; , \\
 \quad m_{\tau} & \simeq & m_{b} \; ,
\end{eqnarray}
also require $M_{e,d}$ being non-diagonal, leading to corrections to the TBM pattern~\cite{Chen:2007afa}.  Note that the complex coefficients in the above mass matrices arise {\it entirely} from the CG coefficients of the $T^{\prime}$ group theory. More precisely, these complex CG coefficients appear in couplings that involve the doublet representations of $T^{\prime}$.

The mass matrices $M_{u}$, $M_{d}$ and $M_{e}$ are diagonalized by,
\begin{eqnarray}
V_{u, R}^{\dagger} M_{u} V_{u,L}  & = &  \mbox{diag} (m_{u}, m_{c}, m_{t}) \; , \\
V_{d, R}^{\dagger} M_{d} V_{d,L}  & = &  \mbox{diag} (m_{d}, m_{s}, m_{b}) \; , \\
V_{e,R}^{\dagger} M_{e} V_{e,L} & = & \mbox{diag} (m_{e}, m_{\mu}, m_{\tau}) \; ,
\end{eqnarray}
where the mass eigenvalues on the right-hand side of the equations are real and positive. 
This gives the following weak charged current interaction in the mass eigenstates of the fermions,
\begin{eqnarray}\label{eq:Lcc}
\mathcal{L}_{cc} & = & \frac{g}{2\sqrt{2}} \biggl[ W^{\mu}_{+}(\vec{x},t) J_{\mu}^{-} (\vec{x},t) 
 + W^{\mu}_{-} (\vec{x},t) J_{\mu}^{+}(\vec{x},t) \biggr] \; , \nonumber \\
J_{\mu}^{-} & = & 
(\overline{u}^{\prime}, \overline{c}^{\prime}, \overline{t}^{\prime})_{L} \gamma_{\mu} V_{CKM}
\left( \begin{array}{c}
d^{\prime} \\
s^{\prime} \\
b^{\prime}
\end{array}\right)_{L} \; . 
\end{eqnarray}
The complex mass matrices $M_{u,d}$ lead to a complex quark mixing matrix, 
\begin{equation}
V_{CKM} = V_{u,L}^{\dagger} V_{d,L} \; .
\end{equation}
The relation
\begin{equation}\label{eq:cabibbo}
\theta_{c} \simeq \biggl| \sqrt{\frac{m_{d}}{m_{s}}} - \sqrt{\frac{m_{u}}{m_{c}}} \biggr|  \simeq \sqrt{\frac{m_{d}}{m_{s}}} \; 
\end{equation}
  is manifest in our model.
 Similarly, the mixing angle $\theta_{12}^{e}$ in the diagonalization matrix $V_{e,L}$ for the charged lepton sector is given by,
 \begin{equation}
 \theta_{12}^{e} \simeq  \sqrt{\frac{m_{e}}{m_{\mu}}} \; . 
 \end{equation}
 Using the Georgi-Jarlskog relations, one then obtains the following relation between the Cabibbo angle and the mixing angle $\theta_{12}^{e}$ in the charged lepton sector,
 \begin{equation}
 \theta_{12}^{e} \simeq \frac{1}{3} \theta_{c} \; .
 \end{equation}
All other elements in $V_{e,L}$ are higher order in $\theta_{c}$, and hence $\theta_{12}^{e}$ gives the dominant corrections to the TBM mixing pattern.  

Due to the discrete symmetries in our model,  the mass hierarchy arises dynamically without invoking an additional U(1) symmetry. The $Z_{12}$ symmetry also forbids Higgsino-mediated proton decays in SUSY version of the model. Due to the $T^{\prime}$ transformation property of the matter fields, the $b$-quark mass can be generated only when the $T^{\prime}$ symmetry is broken, which naturally explains  the hierarchy between $m_{b}$ and $m_{t}$. 
The $Z_{12} \times Z_{12}^{\prime}$ symmetry, to a very high order, also forbids operators that lead to nucleon decays. In principle, a symmetry smaller than  $Z_{12} \times Z_{12}^{\prime}$ would suffice in getting realistic masses and mixing pattern; however, more operators will be allowed and the model would not be as predictive.  

The interactions in $\mathcal{L}_{FF}$ lead to the following neutrino mass matrix, 
\begin{equation}\label{eq:fd}
M_{\nu} = \left( \begin{array}{ccc}
2\xi_{0} + u_{0} & -\xi_{0} & -\xi_{0} \\
-\xi_{0} & 2\xi_{0} & -\xi_{0} + u_{0} \\
-\xi_{0} & -\xi_{0} + u_{0} & 2\xi_{0} 
\end{array} \right) \frac{\lambda v^{2}}{M_{x}} \; ,
\end{equation}
which is parametrized by {\it two} parameters, giving the three absolute neutrino masses~\cite{Chen:2007afa}. As these interactions involve only the triplet representations of $T^{\prime}$, the relevant product rule is $3 \otimes 3$. Consequently, all CG coefficients are real, leading to a real neutrino Majorana mass matrix. The neutrino mass matrix given in Eq.~\ref{eq:fd} has the special property that it is form diagonalizable, {\it i.e.} independent of the values of $\xi_{0}$ and $u_{0}$, it is diagonalized by the tri-bimaximal mixing matrix,
\begin{eqnarray}
U_{\mbox{\tiny TBM}}^{T} M_{\nu} U_{\mbox{\tiny TBM}}  & = &   \mbox{diag}(u_{0} + 3 \xi_{0}, u_{0}, -u_{0}+3\xi_{0}) \frac{v_{u}^{2}}{M_{X}}  \; , \nonumber \\
 & \equiv &   \mbox{diag} (m_{1}, m_{2}, m_{3})   \; .
\end{eqnarray}
While the neutrino mass matrix is real, the complex charged lepton mass matrix $M_{e}$, leads to a complex 
\begin{equation}
V_{\mbox{\tiny PMNS}} = V_{e, L}^{\dagger} U_{\mbox{\tiny TBM}} \; .
\end{equation}
 The Georgi-Jarlskog relations for three generations are obtained. This inevitably requires non-vanishing mixing in the charged lepton sector, as mentioned previously, leading to corrections to the tri-bimaximal mixing pattern.  Consequently, our model predicts a non-vanishing $\theta_{13}$, which is related to the Cabibbo angle as, 
\begin{equation}
\theta_{13}\sim \theta_{c}/3\sqrt{2}\; .
\end{equation}  
Numerically, this is close to $\sin\theta_{13} \sim 0.05$ which is accessible to the Daya Bay reactor experiment. In addition, our model gives rise to a sum rule between the Cabibbo and the solar mixing angle for the neutrinos, 
\begin{equation}
\tan^{2}\theta_{\odot} \simeq \tan^{2} \theta_{\odot, \mathrm{TBM}} + \frac{1}{2} \theta_{c} \cos\delta \; ,
\end{equation} 
which is a consequence of the Georgi-Jarlskog relations in the quark sector. Here the parameter $\delta$ is the Dirac CP phase in the lepton sector in the standard parametrization. This deviation could account for the difference between the experimental best fit value for the solar mixing angle and the value predicted by the tri-bimaximal mixing matrix. 

Since the three absolute neutrino mass eigenvalues are determined by only two parameters, {\it i.e.} the VEVs $u_{0}$ and $\xi_{0}$, there is a sum rule that relates the three light masses, 
\begin{equation}
m_{1} - m_{3} = 2m_{2} \; .
\end{equation}
These masses also satisfy the following sum rule,
\begin{equation}\label{eq:normal}
\Delta m_{\odot}^{2} = -9 \xi_{0}^{2} + \frac{1}{2} \Delta m_{atm}^{2} \; .
\end{equation}
Given that $\Delta m_{\odot}^{2} > 0$ is required in order to have matter effects in solar neutrino oscillation, it immediately follows from the above sum rule given in Eq.~\ref{eq:normal}  that the normal hierarchy pattern with $\Delta m_{atm}^{2} > 0$ is predicted~\cite{Chen:2008eq}.

\section{Numerical Results}

The predicted charged fermion mass matrices in our model are parametrized in terms of 7 parameters~\cite{Chen:2007afa},
\begin{eqnarray}
\frac{M_{u}}{y_{t} v_{u}} & = & \left( \begin{array}{ccccc}
i g & ~~ &  \frac{1-i}{2}  g & ~~ & 0\\
\frac{1-i}{2} g & & g + (1-\frac{i}{2}) h  & & k\\
0 & & k & & 1
\end{array}\right)  , \\
\frac{M_{d}, \; M_{e}^{T}}{y_{b} v_{d} \phi_{0}\zeta_{0}}  & = &  \left( \begin{array}{ccccc}
0 & ~~ & (1+i) b & ~~ & 0\\
-(1-i) b & & (1,-3) c & & 0\\
b & &b & & 1
\end{array}\right)  \; .
\end{eqnarray}
With the input parameters
\begin{eqnarray}
b \equiv \phi_{0} \psi^{\prime}_{0} /\zeta_{0} = 0.00304 \; , \; c\equiv \psi_{0}N_{0}/\zeta_{0}=-0.0172 \; , \\
 k \equiv y^{\prime}\psi_{0}\zeta_{0}=-0.0266 \; , \; 
h\equiv \phi_{0}^{2}=0.00426 \; , \; g \equiv \phi_{0}^{\prime 3}= 1.45\times 10^{-5} \; ,
\end{eqnarray}
the following mass ratios are obtained,
\begin{eqnarray}
m_{d}: m_{s} : m_{b} & \simeq & \theta_{c}^{\scriptscriptstyle 4.7} : \theta_{c}^{\scriptscriptstyle 2.7} : 1 \; , \\ 
m_{u} : m_{c} : m_{t} & \simeq &  \theta_{c}^{\scriptscriptstyle 7.5} : \theta_{c}^{\scriptscriptstyle 3.7} : 1 \; ,
\end{eqnarray} 
with $\theta_{c} \simeq \sqrt{m_{d}/m_{s}} \simeq 0.225$. We have also taken $y_{t} = 1.25$ and $y_{b}\phi_{0} \zeta_{0} \simeq m_{b}/m_{t} \simeq 0.011$ and have taken into account the renormalization group corrections. As a result  of the Georgi-Jarlskog relations, realistic charged lepton masses are obtained. 
These parameters also gives rise to the following complex CKM matrix,
\begin{eqnarray}
\left( \begin{array}{ccc}
0.974e^{-i 25.4^{o}} & 0.227 e^{i23.1^{o}} & 0.00412e^{i166^{o}} \\
0.227 e^{i123^{o}} & 0.973 e^{-i8.24^{o}} & 0.0412 e^{i180^{o}} \\
0.00718 e^{i99.7^{o}} & 0.0408 e^{-i7.28^{o}} & 0.999
\end{array}\right). 
\end{eqnarray}
The predictions of our model for the angles in the unitarity triangle  and the Jarlskog invariant in the quark sector are,
\begin{eqnarray}
\beta & \equiv &  \mbox{arg} \biggl( \frac{-V_{cd} V_{cb}^{\ast}}{V_{td}V_{tb}^{\ast}} \biggr) = 23.6^{o}, \; \sin2\beta  =  0.734 \; ,  
\\
\alpha  & \equiv &  \mbox{arg} \biggl( \frac{-V_{td} V_{tb}^{\ast}}{V_{ud}V_{ub}^{\ast}} \biggr) = 110^{o} \; , \\
\gamma  & \equiv &  \mbox{arg} \biggl( \frac{-V_{ud} V_{ub}^{\ast}}{V_{cd}V_{cb}^{\ast}} \biggr)  = \delta_{q} = 45.6^{o} \; , \\
J  & \equiv &  \mbox{Im} (V_{ud} V_{cb} V_{ub}^{\ast} V_{cd}^{\ast}) = 2.69 \times 10^{-5} \; ,
\end{eqnarray}
where $\delta_{q}$ is the CP phase in the standard parametrization, which has a large experimental uncertainty at present. 
In terms of the Wolfenstein parameters, we have 
\begin{eqnarray}
\lambda & = & 0.227 \; , \\
A & = & 0.798 \; , \\
\overline{\rho} & = & 0.299 \; , \\
\overline{\eta} & = & 0.306 \; .
\end{eqnarray}

We compare our predictions to the experimental values from CKMFitter collaboration reported at Moriond 2009~\cite{moriond09,Amsler:2008zzb}. 
The $3\sigma$ allowed range for the CKM matrix elements given by
\begin{eqnarray}
|V_{ud}| & = & 0.9737-0.9749 \\
|V_{us}| & = & 0.2227 - 0.2277 \\
|V_{ub}| & = & 0.00321-0.00394 \\
|V_{cd}| & = & 0.2226-0.2276 \\
|V_{cs}| & = & 0.9729-0.9741 \\
|V_{cb}| & = & 0.0393-0.0423  \\
|V_{td}| & = & 0.00795-0.00915 \\
|V_{ts}| & = & 0.0385-0.0415 \\
|V_{tb}| & = & 0.9991-0.9992
\end{eqnarray}
For the three angles of the unitarity triangle, the limits from {\it direct} measurements~\cite{moriond09,Amsler:2008zzb} at $3\sigma$ are, 
\begin{eqnarray}
\alpha & = & 76^{o}- 110^{o}  \; , \\
\beta & = & 20.1^{o}-30.2^{o} \; , \\
\gamma & = & 18^{o}-130^{o}  \; .
\end{eqnarray}
And the $3\sigma$ limits for the Wolfenstein parameters are given by, 
\begin{eqnarray}
A & = & 0.767-0.841 \; , \\
\lambda & = & 0.2227-0.2277  \; ,\\
\overline{\rho} & = & 0.087-0.212  \; , \\
\overline{\eta} & = & 0.307-0.389 \; , \\
J & = & (2.69-3.37) \times 10^{-5} \; .
\end{eqnarray}
We note that except for $\overline{\rho}$ and $|V_{td}|$, all other parameters are consistent with the current $3\sigma$ experimental limits. Our prediction for $\overline{\rho}$ is slightly higher than the experimental upper bound while the prediction for $|V_{td}|$ is slightly lower than the experimental lower limit. Given the tension that is currently present in the global fit between $\sin2\beta$ and $|V_{ub}|$ (which is also the fit that gives the experimental value for $\overline{\rho}$), and the fact that there are still large theoretical uncertainty in hadronic effects when extracting the experimental value for $|V_{td}|$, the experimental determination for the values of $\overline{\rho}$ and $|V_{td}|$ thus have some uncertainty at present. 

In the lepton sector,  the diagonalization matrix for the charged lepton mass matrix combined with $U_{TBM}$ gives numerically the following PMNS matrix, 
\begin{equation}
\left( \begin{array}{ccc}
0.838 e^{-i178^{o}} & 0.543 e^{-i173^{o}} & 0.0582 e^{i138^{o}}  \\
0.362 e^{-i3.99^{o}} & 0.610 e^{-i173^{o}} & 0.705 e^{i3.55^{o}}  \\
0.408 e^{i180^{o}} & 0.577 & 0.707
\end{array}\right) \; ,
\end{equation}
which predicts 
\begin{eqnarray}
\sin^{2}\theta_{\mathrm{atm}} & = & 1 \; , \\
\tan^{2}\theta_{\odot} & = & 0.420 \; , \\
|U_{e3}| & = & 0.0583 \; .
\end{eqnarray}
The two VEV's, 
\begin{equation}
u_{0} = -0.0593 \quad , \quad \xi_{0} = 0.0369
\end{equation}
 give 
 \begin{eqnarray}
 \Delta m_{atm}^{2} & = & 2.4 \times 10^{-3} \; \mbox{eV}^{2} \; , \\
 \Delta m_{\odot}^{2} & = & 8.0 \times 10^{-5} \; \mbox{eV}^{2} \; .
 \end{eqnarray}
 The leptonic Jarlskog is predicted to be 
 \begin{equation}
 J_{\ell} = -0.00967 \; ,
 \end{equation}
 and equivalently, this gives  a Dirac CP phase, 
 \begin{equation}
 \delta_{\ell} = 227^{o} \; . 
 \end{equation}
 With such $\delta_{\ell}$, the correction from the charged lepton sector can account for the difference between the TBM prediction and  the current best fit value for $\theta_{\odot}$. Our model predicts 
\begin{eqnarray}
m_{1} & = & 0.0156 \; \mbox{eV} \; , \\
m_{2} & = & -0.0179 \; \mbox{eV} \; , \\
m_{3} & = & 0.0514 \;  \mbox{eV} \; , 
\end{eqnarray}
with Majorana phases 
\begin{eqnarray}
\alpha_{21} & = & \pi \; , \\
\alpha_{31} & = & 0 \; . 
\end{eqnarray}

Since the leptonic Dirac CP phase, $\delta_{\ell}$, is the only non-vanishing CP violating phase in the lepton sector, a connection~\cite{Chen:2007fv} between leptogenesis and low energy CP violating leptonic processes, such as neutrino oscillation, can exist in our model.

\section{Conclusion}

We present a model based on $SU(5)$ and the double tetrahedral group $T^{\prime}$ as the family symmetry. CP violation in our model is entirely geometrical due to the presence of the complex group theoretical CG coefficients in $T^{\prime}$. The Georgi-Jarlskog relations automatically lead to a sum rule between the Cabibbo angle and the solar mixing angle for the neutrino. The predicted CP violation measures in the quark sector are consistent with the current experimental data. The leptonic Dirac CP violating phase is predicted ~\cite{Chen:2009gf} to be $\delta_{\ell} \sim 227^{o}$, which gives the cosmological matter-antimatter asymmetry~\cite{Chen:2009}.

\section*{Acknowledgments}
The work of M-CC was supported, in part, by the National Science Foundation under grant no. PHY-0709742. The work of KTM was supported, in part, by the Department of Energy under Grant no. DE-FG02-04ER41290.

\end{document}